**Article** **Open Access**

# The FAST Core Array

Peng Jiang[1,2]*, Rurong Chen[1,2], Hengqian Gan[1,2], Jinghai Sun[1,2], Boqin Zhu[1,2], Hui Li[1,2], Weiwei Zhu[1], Jingwen Wu[1], Xuelei Chen[1], Haiyan Zhang[1,2], Tao An[3]

[1]*National Astronomical Observatories*, *Chinese Academy of Sciences*, *Beijing* 100101, *China*

[2]*Guizhou Radio Astronomical Observatory*, *Guizhou University*, *Guiyang* 550000, *China*

[3]*Shanghai Astronomical Observatory*, *Chinese Academy of Sciences*, *Shanghai* 200030, *China*

*Correspondence: pjiang@nao.cas.cn

**Abstract:** The Five-hundred-meter Aperture Spherical Radio Telescope (FAST) Core Array is a proposed extension of FAST, integrating 24 secondary 40-m antennas implanted within 5 km of the FAST site. This original array design will combine the unprecedented sensitivity of FAST with a high angular resolution (4.3" at a frequency of 1.4 GHz), thereby exceeding the capabilities at similar frequencies of next-generation arrays such as the Square Kilometre Array Phase 1 or the next-generation Very Large Array. This article presents the technical specifications of the FAST Core Array, evaluates its potential relatively to existing radio telescope arrays, and describes its expected scientific prospects. The proposed array will be equipped with technologically advanced backend devices, such as real-time signal processing systems. A phased array feed receiver will be mounted on FAST to improve the survey efficiency of the FAST Core Array, whose broad frequency coverage and large field of view（FOV） will be essential to study transient cosmic phenomena such as fast radio bursts and gravitational wave events, to conduct surveys and resolve structures in neutral hydrogen galaxies, to monitor or detect pulsars, and to investigate exoplanetary systems. Finally, the FAST Core Array can strengthen China's major role in the global radio astronomy community, owing to a wide range of potential scientific applications from cosmology to exoplanet science.

## 1. INTRODUCTION

The Five-hundred-meter Aperture Spherical Radio Telescope (FAST), completed and operational since February 2020, is currently the largest and most sensitive single-dish radio telescope worldwide[1,2]. Its considerable diameter and favorable location in a deep karst depression, shielded from anthropogenic radio interferences, have already allowed operating scientists to produce noteworthy scientific discoveries that include new pulsars detection[3,4], fast radio bursts[5–9] and microquasars[10] investigations, and analyses of neutral atomic hydrogen (HI) and interstellar molecules[11,12]. The sensitivity of FAST at low frequencies exceeds that of the former 305-m Arecibo telescope, damaged in 2020 and subsequently decommissioned.

However, several next-generation radio interferometric arrays will be completed by 2035, representing increasing international competition for FAST. Completion of the Square Kilometre Array Phase 1 (SKA1, Australia–South Africa) is scheduled for 2029[13], and that of the next-generation Very Large Array (ngVLA, United States of America) by 2035[14]. Both arrays will provide considerably higher sensitivity, better angular resolution, and faster survey speed than existing instruments. Although FAST will still offer the highest sensitivity, its spatial resolution, constrained by its fixed single aperture, will be exceeded by the SKA1 and ngVLA. Moreover, FAST is a very large single-dish telescope, thus calibration, accurate imaging of complex structures, and source localization themselves are challenging.

To expand FAST capabilities by its integration into a synthesis array, the FAST Core Array was proposed by the FAST Operation and Development Center. The main objectives of the FAST Core Array are to maintain China's competitiveness among next-generation radio astronomy facilities through the 2030s and to provide a proof-of-concept for the future FAST Array (FASTA) to acquire invaluable technical experience and anticipate potential problems during FASTA implementation[15]. Furthermore, once FASTA is completed, the FAST Core Array antennas can complement FASTA with shorter baselines to improve the imaging quality. Finally, the FAST Core Array could also be applied to follow-up studies of FASTA scientific discoveries.

The radio quiet zone with a 30 km radius centered on FAST, located in Guizhou Province, is a major strategic advantage for the FAST Core Array. Radio frequency protection regulations have been implemented[16] to minimize electromagnetic interference from human activities during sensitive astronomical observations. In addition, optimization of flight paths over FAST effectively prevents radio frequency interference from aircraft[17]. This high-quality radio environment was achieved through substantial financial investment and cross-sector collaboration. Constructing the FAST Core Array in this protected zone will provide nearly ideal electromagnetic observation conditions, while stimulating local economic development around the FAST site. By maximizing the use of existing infrastructure, the Core Array will take full advantage of FAST sensitivity in an optimized radio environment.

The additional antennas will function as a synthesis aperture array when combined with FAST. Unfortunately, because of factors inherent to interferometric imaging and signal correlation, synthesis aperture arrays frequently suffer image sensitivity losses in comparison with single-dish telescopes. For interferometric imaging, the number of antennas required to ensure a FAST Core Array image sensitivity comparable with FAST's single-dish mode (antenna efficiency $A_{\text{eff}}$ = 0.65, system temperature $T_{\text{sys}}$ = 20 K[18]) can be estimated from their aperture. For example, maintaining the sensitivity level would require 4, 8, 16, 24, and 44 antennas with diameters of 100 m, 70 m, 50 m, 40 m, and 30 m, respectively. Cost scaling as a function of antenna size is an important consideration, because the cost per unit collecting area increases nonlinearly when telescope diameters exceed 40 m. Specifically, the construction cost of a 40-m antenna is approximately ten million RMB, whereas that of a 50-m antenna is approximately four times higher. Moreover, the cost of backend equipment depends linearly on the number of antennas, while the cost of data storage and processing depends on that number squared. Additionally, the beam size of a 40-m aperture antenna is comparable with the FOV of FAST's current 19-beam receiver. Finally, the complex karst landscape surrounding FAST does not provide sufficient space to accommodate numerous small antennas, unlike the SKA1 or ngVLA sites. Therefore, considering cost constraints, sensitivity requirements, terrain limitations, and necessary integration with FAST, a 40-m aperture was selected as the optimal design for the FAST Core Array antennas.

The expected sensitivity of the FAST Core Array will be approximately 3 000 m²/K (40-m antennas, $A_{\text{eff}}$ = 0.7, $T_{\text{sys}}$ = 30 K), nearly twice that of the mid-frequency SKA1 component (1 560 m²/K, $A_{\text{eff}}$ = 0.9, $T_{\text{sys}}$ = 18 K[13,19]) and comparable to the ngVLA specifications (3 050 m²/K, $A_{\text{eff}}$ = 0.828, $T_{\text{sys}}$ = 17 K[20]). Achieving this efficiency will ensure strong technological competitiveness.

The following sections describe the technical array specifications, scientific motivation, key technological components (antenna design, site selection, array design, Phased Array Feed (PAF) receivers, and control and data centers), and scientific research objectives of the FAST Core Array.

## 2. SCIENTIFIC MOTIVATION AND TECHNICAL SPECIFICATIONS

### 2.1. Scientific Motivation

The primary scientific motivation for the FAST Core Array is to expand FAST's capabilities for high-resolution radio image acquisition without compromising on its exceptional sensitivity, thereby improving image quality for numerous celestial sources and expanding the scientific scope of possible research beyond that of FAST alone. Key science objectives include HI galaxies, continuum emission from radio galaxies, pulsars, transient radio phenomena such as fast radio bursts (FRBs) or gamma-ray bursts, and radio counterparts of gravitational wave events. Additionally, the FAST Core Array will acquire spectroscopic observations for enhanced analysis of phenomena involving interstellar molecules, such as hydroxyl masers (Section 4).

### 2.2. Technical Specifications

The FAST Core Array combines FAST with 24 additional 40-m diameter antennas positioned within a 5-km radius from FAST. The first phase aims to deploy six antennas by 2026. In this first-phase configuration, the achieved angular resolution will be 4.3" at a frequency of 1.4 GHz. The full-scale FAST Core Array (Table 1) is expected by 2030. FAST is currently equipped with a 19-beam receiver[18] with an FOV comparable to the beam size (18' at 1.4 GHz) of a 40-m antenna. The 40-m antenna pointing accuracy is approximately 15", with a minimum elevation of 5°. In comparison, FAST's pointing accuracy is 8", with a minimum elevation of 50°. Therefore, the overall sky coverage achievable with the FAST Core Array will be primarily determined by the fraction of the sky accessible to FAST. By combining FAST with 24 secondary 40-m antennas, the synthesized array sensitivity and angular resolution will markedly improve on FAST's single-dish performance.

The expected efficiency of the FAST Core Array antennas is approximately 0.7, with a system temperature of 30 K in the L band. In comparison, antenna efficiency for the central beam of FAST's current 19-beam receiver is 0.65, with a system temperature of 20 K. With 24 secondary antennas combined with FAST, the overall sensitivity of the FAST Core Array ($A_{\text{eff}}/T_{\text{sys}}$) should reach 3 000 m²/K. Even without FAST, the sensitivity of the array comprising only the 24 secondary antennas would be 703 m²/K, approximately 3.5 times that of the Karl G. Jansky VLA[21] and 1.5 times that of the Meer Karoo Array Telescope (MeerKAT[19]).

**Table 1. Key technical specifications of the FAST Core Array**

| | |
|---|---|
| $A_{\text{eff}}/T_{\text{sys}}$ (m²/K @1.4 GHz) | ~ 3 000 |
| Frequency range/GHz | 0.35–10 |
| Max baseline/km | ~ 10 |
| Resolution (arcsec @1.4 GHz) | ~ 4.3 |
| FOV (arcmin @1.4 GHz) | 18 |

## 3. KEY TECHNOLOGICAL COMPONENTS

### 3.1. Secondary Antenna Design

Each 40-m antenna (Table 2) follows a primary focus design with a focal-to-diameter ratio of 0.375. The entire antenna assembly weighs approximately 400 tons. It consists of three primary subsystems: the antenna feed subsystem, structural subsystem, and servo subsystem. The antenna schematic structure is illustrated in Fig. 1.

The antenna feed subsystem comprises the antenna reflector and the feed cabin. The central 20-m region of the main reflector is plated with solid aluminum surface panels. The outer region is covered with stainless-steel wire mesh surface panels supported by aluminum ribs. The main reflector is divided into 11 concentric zones, reinforced with stainless-steel wire mesh and aluminum ribs and assembled from sets of 12 and 24 panels for the innermost zones and 48 panels for each of the remaining nine zones, representing a total of 468 panels. With accessible elevations within 5°–90°, surface shape accuracy of the full aperture for a single antenna is better than 3 mm (rms), within observational requirements at frequencies of up to 10 GHz. Combined implementation of solid and mesh panels in the central and outer regions of the reflector, respectively, is a compromise to achieve the required collecting area while both constraining structural properties and ensuring that the antennas will withstand weather conditions through the operational lifetime of the FAST Core Array.

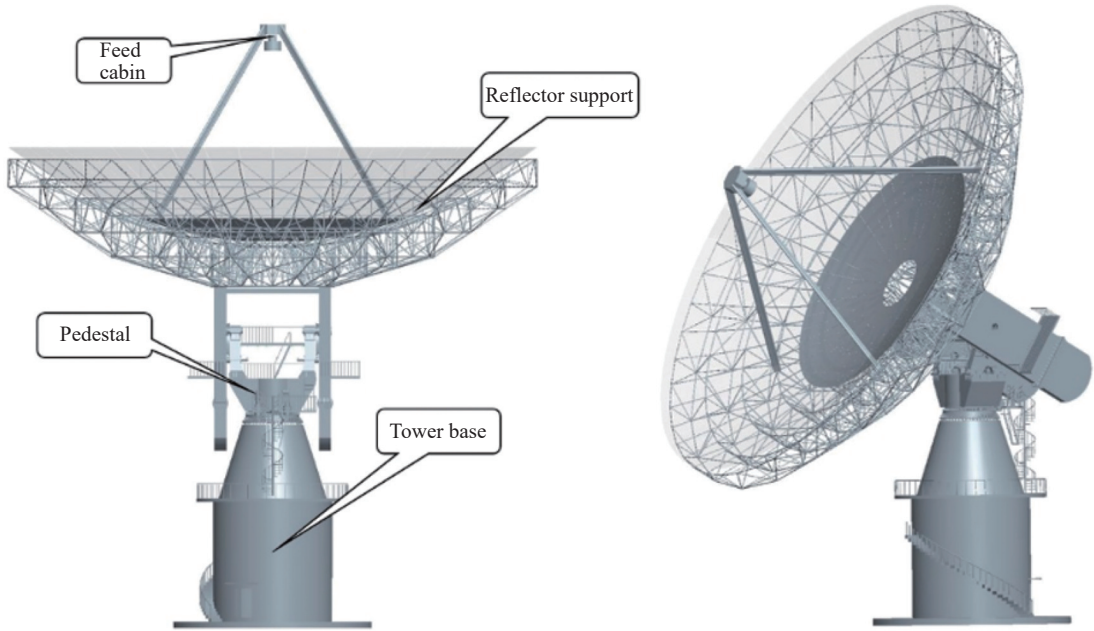

**Fig. 1. Schematic structural diagram of a 40-m antenna to be integrated in the FAST Core Array.**

**Table 2. Main technical specifications of a single antenna**

| | |
|---:|:---:|
| Aperture/m | 40 |
| $f/D$ ratio | 0.375 |
| Resolution (arcmin @21 cm) | 18 |
| Frequency range/GHz | 0.35–10 |
| $A_{\text{eff}}/T_{\text{sys}}$ (m²/K @L band) | 29.3 |
| Pointing/(") | 15 |
| Min elevation/(°) | 5 |

The feed cabin is supported by four columns located at the focal point of the main reflector. Its dimensions are 4 m × 4 m × 1.6 m. It houses a feed turntable, on which several multiple-band feeds and receivers are installed (Table 3).

Each antenna will be equipped with three sets of customizable feeds (Fig. 2) to optimize sensitivity across different frequency ranges. Each feed provides illumination within an angle of approximately 140°, with associated low-noise amplifiers operating at room temperature. Besides the feeds and low-noise amplifiers, the receivers include 90° couplers, filters, isolators, amplifier groups, mixers, and analog-to-digital conversion modules. First, dual-linearly polarized signals from the feeds are con-

86    www.ati.ac.cn

## Table 3. Feed technical specifications for a single antenna

| Frequency/GHz | Feed form | Feed size/cm | Cryogenic | $T_{sys}$/K |
| --- | --- | --- | --- | --- |
| 0.35–1.05 | Circular eleven | 55×19 | No | 50–150 |
| 1.1–3.3 | Coaxial corrugated horn | 65×150 | No | 30 |
| 3.3–10 | Quad-ridge flared horn | 60×100 | No | 35 |

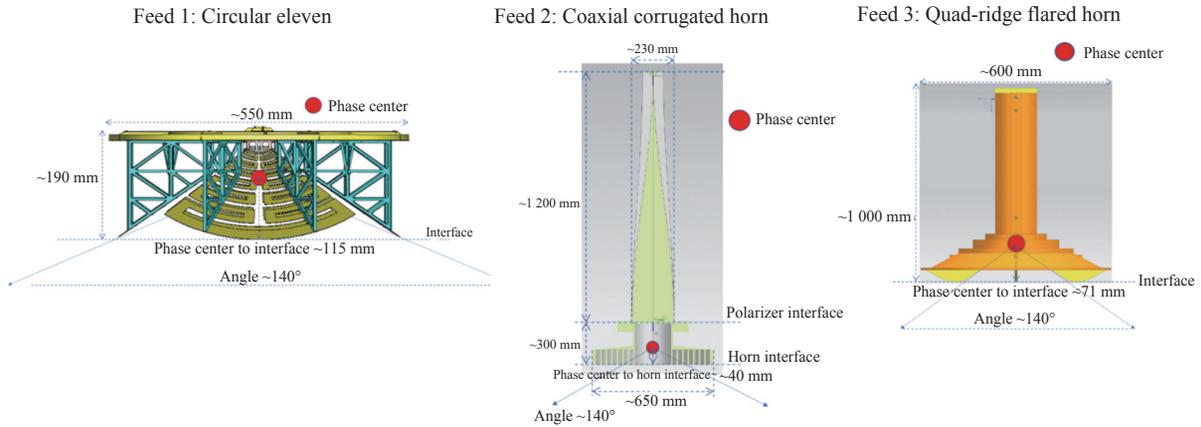

**Fig. 2.** Preliminary design diagrams for three single-antenna tentative feeds. Phase centers are indicated by filled red circles.

verted to dual-circular polarizations with a 90° coupler, and then digitized. The digitized signals are then transmitted by optical fiber to the array data center for correlation, calibration, imaging, and analysis.

The structural subsystem comprises the main reflector support, feed cabin support, antenna pedestal, and tower base. The main reflector support consists of 24 primary and 24 secondary radial beams, with additional radial and diagonal struts. Feed cabin support uses a rectangular beam structure with four legs distributed in a plane inclined at 45° from the elevation surface. The antenna pedestal uses an azimuth-elevation turntable for pointing, with an azimuthal operational range of ±270° and an elevation operational range of 5°–90°. Azimuthal angular velocity is within requirements (0.003°/s–1°/s), with a maximum angular acceleration of 0.3°/s². Elevation angular velocity is also within requirements (0.001°/s–1°/s), with a maximum angular acceleration of 0.3°/s².

The servo subsystem is designed for antenna steering and control. It comprises an antenna control unit, a drive control unit, drive motors, an axis-angle encoder unit, and a safety protection unit. During operation, the status of each antenna can be monitored and controlled remotely with the central control system located in the telescope control center (Section 3.4).

Finally, a dedicated shielded control room, associated with each antenna, will house a local antenna control unit, a time-frequency unit, and the local digital backend.

### 3.2. Site Selection and Array Design

The area surrounding FAST is characterized by a complex karst topography. Therefore, preliminary site selection for the FAST Core Array must simultaneously account for several major geological, infrastructural, and observational constraints. Specifically, locations convenient for construction logistics and with assured access to essential utilities such as water, electricity, and transportation were prioritized. Additionally, potential topographic obstructions were considered and avoided to reduce limitations on antenna elevation and to increase observational sky coverage. Within these constraints, 35 preliminary locations were identified during site selection, with seven potential sites classified as highly suitable (Fig. 3, red circles), sixteen as moderately suitable (Fig. 3, orange circles), and twelve as low-suitability sites (Fig. 3, green circles).

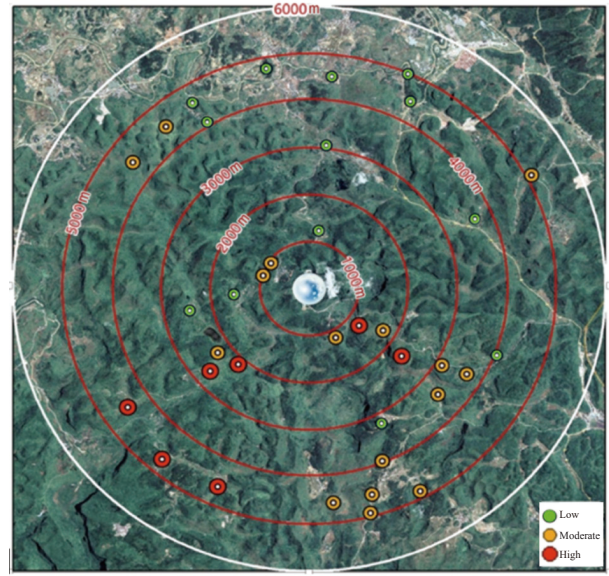

**Fig. 3.** Preliminary site selection for the FAST Core Array. Candidates are located within 5 km (white circle) of FAST. They are classified by suitability: high (red filled circles), moderate (orange), and low (green).

The design of the FAST Core Array interferometric configuration was carefully determined, primarily on the basis of scientific objectives and planned observation modes, to maximize imaging quality and efficiency. Because primary observational targets of the FAST Core Array include celestial objects with low surface brightness, optimizing the array design is critical to conduct efficient surveys, particularly for short baselines (within 5 km of the FAST site).

In this context, a Reuleaux triangle configuration is well-suited to maximize sidelobe suppression. To assess the feasibility of this configuration for the FAST Core Array, 24 of the 35 preliminary locations were selected as potential sites for antenna implantation (Fig. 4).

Subsequently, observation simulations were conducted for a celestial source at a declination of 30°, with acquisition times between 12 min and 4 h. Resulting UV (spatial frequencies in the E–W and N–S directions) coverage and dirty beam are illustrated in Fig. 5 (Columns 1 and 2, respectively). Point-spread-function profiles demonstrated excellent sidelobe suppression in this configuration (Fig. 5, Column 3), to below 5% for a 2-h observation (Fig. 5, Panel $E_3$). This is within requirements for most imaging observations. Moreover, a 12-min simulation also exhibited acceptably uniform UV coverage with sidelobes suppressed at the 10% level, within the requirements for rapid survey observations. In equal bandwidth conditions, this implies that image sensitivity of the FAST Core Array for an acquisition time of 12 min should be an order of magnitude higher than that of the "Faint Images of the Radio Sky at Twenty Centimeters" survey (acquisition time: 3 min) conducted with the VLA[22].

Among the 24 potential sites initially selected, six optimal locations were chosen for the proof-of-concept precursor array (six 40-m antennas). Their geographic distribution around FAST and the resulting baseline lengths are illustrated on the topographic map in Fig. 6 (sites Fea-1 to Fea-6). At the time of writing, final antenna positions have been selected at two sites (Fea-1 and Fea-2, red stars, Fig. 6) where antenna construction is underway.

Scientists of the FAST Core Array team intend to expand the number of potential antenna sites near FAST to fully optimize the final Core Array configuration, if possible, by achieving a final selection of 100 suitable options. Furthermore, data reduction studies are planned as part as the current feasibility research to improve the imaging quality and efficiency of the FAST Core Array.

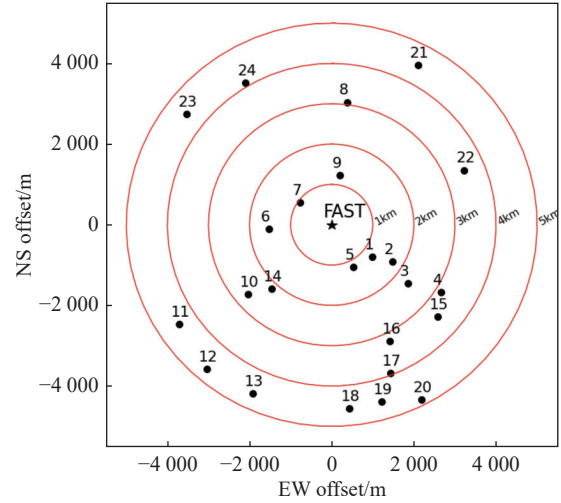

**Fig. 4. Preliminary design for the FAST Core Array, illustrating the positions of the 24 most suitable candidate sites within 5 km of FAST (radial distance indicated by the red concentric circles).**

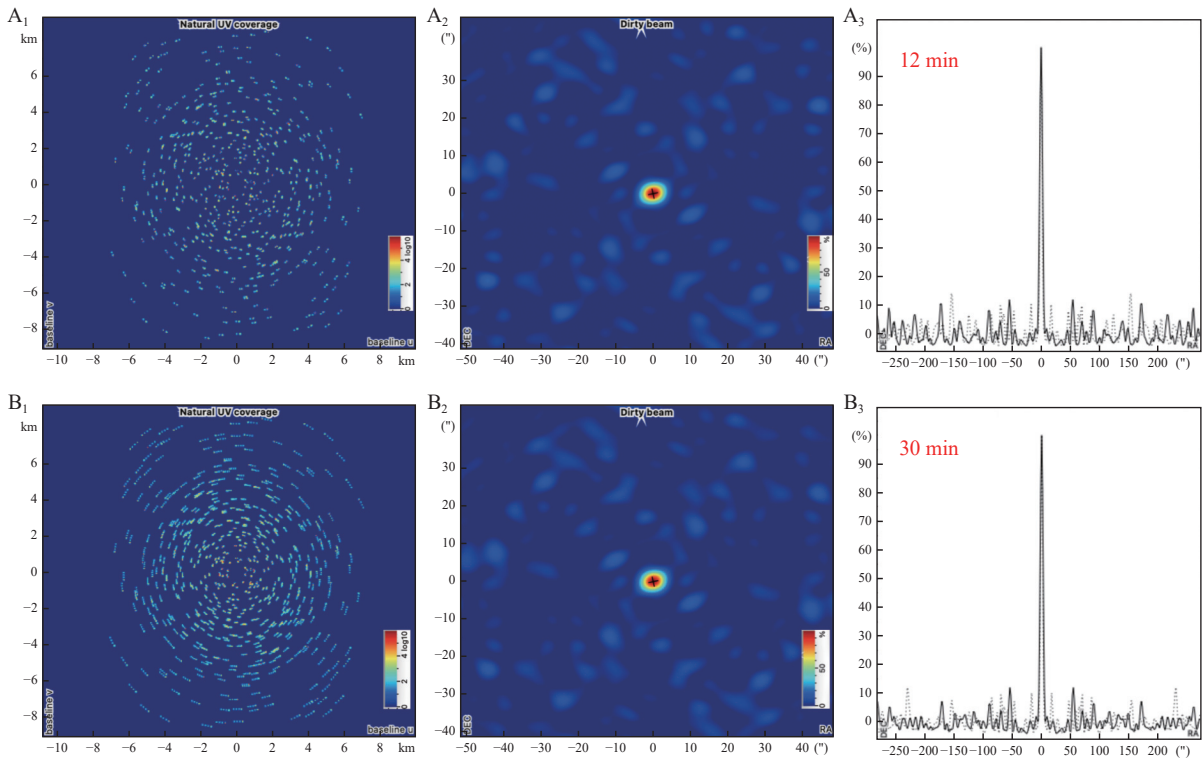

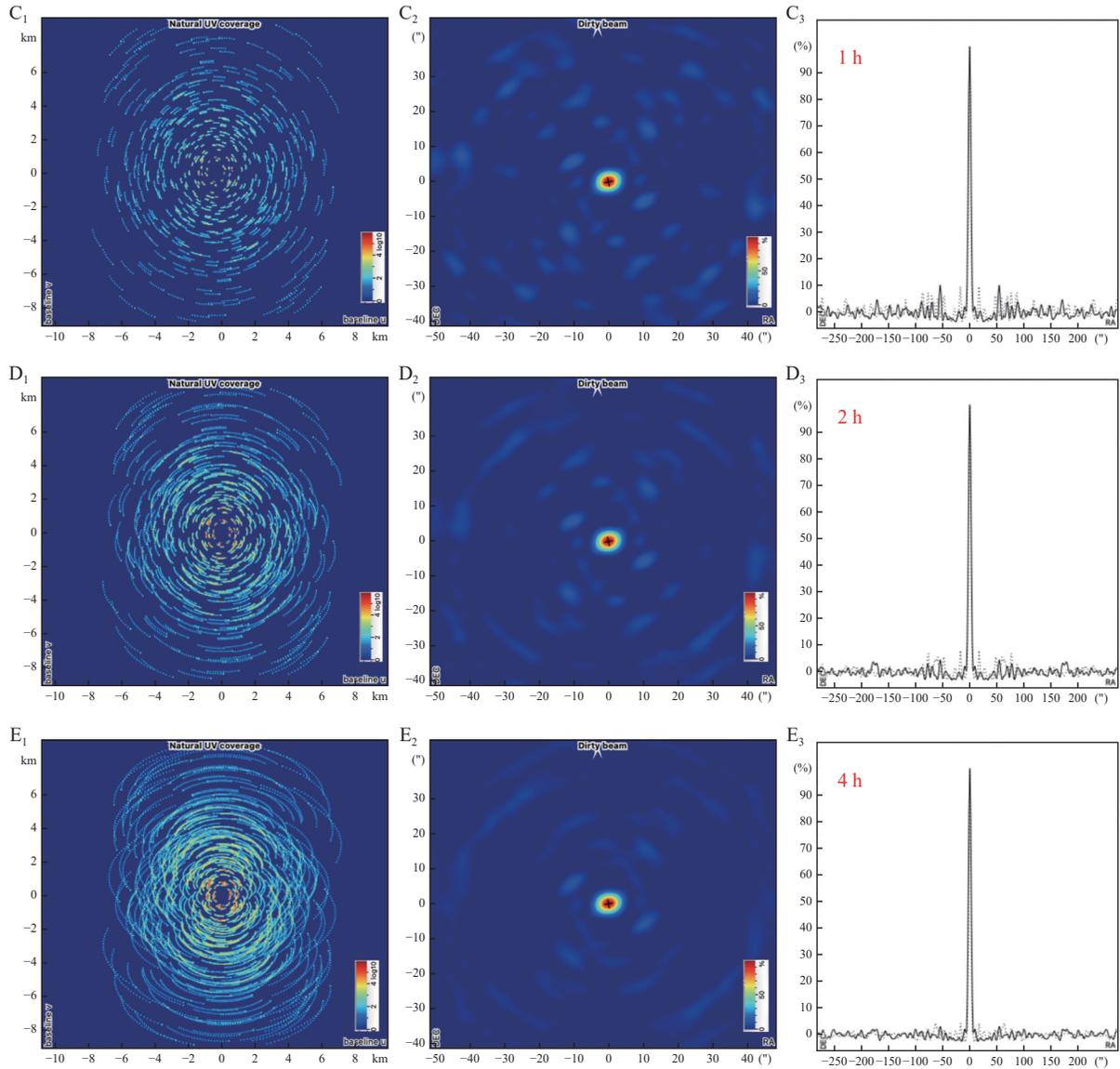

**Fig. 5.** UV coverage (spatial frequencies in the E–W and N–S directions, Column 1), dirty beam (Column 2), and point-spread-function profile (Column 3) of the FAST Core Array for a source at a declination of 30°, with different acquisition times: 12 min, 30 min, 1 h, 2 h, and 4 h (Rows A–E, respectively). The software iAntConfig[23] was used to optimize array configuration and to generate the plots.

### 3.3. PAF Receivers

The aperture of a telescope determines the size of its beam. FAST has an effective aperture of 300 m, resulting in a beam size of 3' at 1.4 GHz. To enhance survey efficiency, FAST is coupled with a 19-beam horn receiver, expanding the FOV to 19 times the nominal single-beam size. However, the frequency range of the horn receiver is narrow (currently 1.05–1.45 GHz), which limits the achievable scientific objectives.

PAF receivers[24] are the next generation of radio receivers, with three main advantages over standard multi-beam horn feed receivers. First, horn feeds are constrained by their physical size, with gaps between adjacent beams requiring multiple observations of the same field to achieve full coverage. Conversely, PAF receivers generate beams in mathematical space, allowing for spatial overlap and providing continuous sampling of the focal plane. Second, PAF receivers achieve frequency ratios of up to 1:3, significantly broadening the available frequency range compared with that of FAST's current 19-beam receiver. Finally, PAF receivers provide a markedly improved illumination function directly related to the main reflector characteristics, yielding higher aperture efficiency.

In addition to the existing 19-beam horn receiver, a wide-field, broadband PAF receiver will be developed for FAST. It will include advanced signal processing technology such as low-loss wideband feeds, new-generation room-temperature amplifiers, digital beamforming algo-rithms, and high-speed digital terminals. The available frequency coverage will increase to 1–3 GHz, with a receiver noise temperature lower than 30 K and a FOV wider than 0.25 (°)$^2$. Implementing the PAF receiver on FAST should enhance its single-dish survey efficiency, diversify its scientific targets, and expand its FOV to a value comparable with that of the 40-m secondary antennas, thereby improving the observation efficiency of the FAST Core Array.

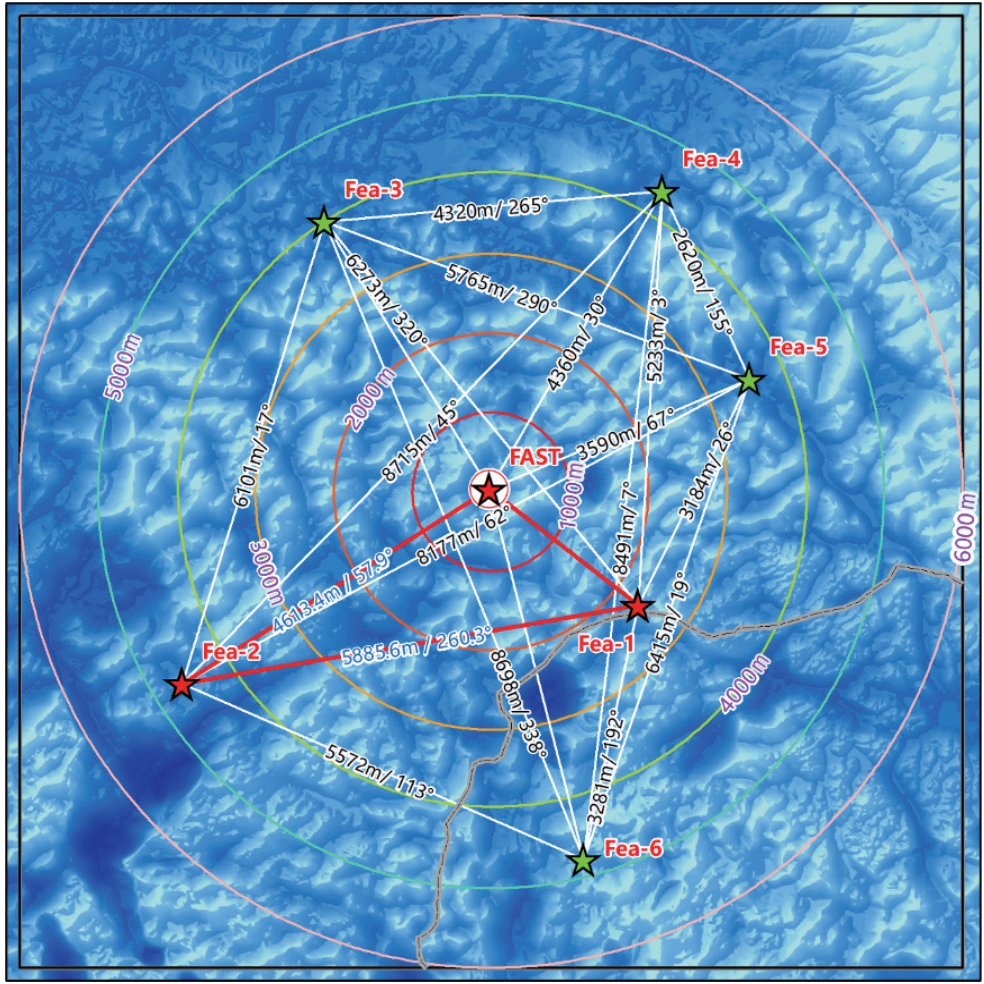

**Fig. 6.** Distribution of the six optimally located antennas for the precursor array (first phase of the FAST Core Array), superimposed on an elevation map around the FAST location (darker colors for low elevations, lighter colors for higher elevations). Positions Fea-1, 2, 3, 4, 5, and 6 correspond to potential sites 2, 11, 24, 21, 22, and 19, respectively, in Fig. 4.

### 3.4. Telescope Control and Data Center

Observational signals collected by each antenna are digitized in the local digital backend. The digitization process occurs locally, enhancing signal resilience to interference during transmission compared with analog signals. Furthermore, the digitized signals are time-stamped at each station before transmission to avoid phase errors during long-distance optical fiber transmission, and then transmitted to the array correlator. Finally, after processing at the correlator center, the data products are archived at the data storage center (Fig. 7).

FAST Core Array facilities include: a telescope control center, equipped with an array monitoring and control system; a time-frequency service system; a correlator center; and a data storage center. Observational data, control data, time-frequency standard signals, and all additional data relevant to interferometric observations are transmitted through optical fibers connecting the control center with each antenna control room. Current high-precision, stable-phase optical fiber technology ensures accurate synchronization of time-frequency signals between the control room and each antenna. Overall array control, as well as diagnostics for the receivers, digital backends, and operation status of the time-frequency system at each station are conducted by the central control system located at the telescope control center.

The correlator is a critical component of the FAST Core Array. Each array antenna generates baseband data streams that necessitate a dedicated hardware correlator for real-time processing. Because of the substantial number of antennas in the array (25 in total) and of computation time and sidelobe minimization constraints, the correlator will operate in the FX mode, in which input signals are first Fourier-transformed (F) then cross-multiplied (X), and perform real-time correlation of the baseband data streams[25].

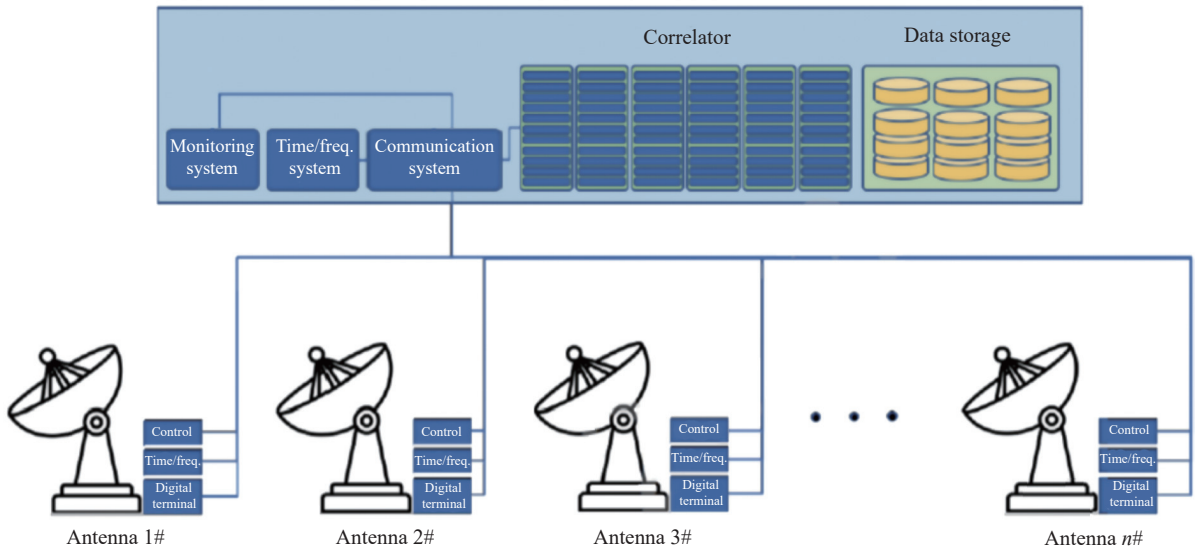

**Fig. 7.** Schematic diagram of the signal processing chain in the FAST Core Array, from the single antenna to the telescope control and data center.

The FAST Core Array correlator operates in three modes: correlation processing, phased array, and Very Long Baseline Interferometry (VLBI). Baseband data streams first enter the F module and undergo delay compensation, Fourier transform, and quantization to produce channelized data streams, which then enter the X module through high-speed switches. On each channel, data from all possible antenna pairs are multiplied and temporally averaged to derive the baseline visibility functions subsequently used for image processing. In the phased array mode, the X module stores the autocorrelation function of each antenna. After delay compensation, the autocorrelation functions are used for phased beamforming, thus effectively considering the FAST Core Array as a larger single-dish telescope, which increases its effective aperture by 42%. This mode is suitable for observations with extreme sensitivity requirements. In the VLBI mode, baseband data streams undergo delay compensation and quantization without Fourier transform. Delayed baseband data are treated as single-station data and combined with baseband data from other stations for correlation processing.

The correlator size will increase as construction progresses. The F module size, determined by the total number of antennas, will use a field-programmable gate array for baseband data stream channelization. The X module size, depending on the number of antennas, bandwidth, and number of channels, will use graphics processing unit (GPU) servers for multiplication and summation operations on the channelized data streams.

## 4. SCIENTIFIC RESEARCH

Implementation of the FAST Core Array is intended to strengthen FAST's position at the forefront of global radio astronomy, by maintaining its extremely high sensitivity while enhancing its high-resolution imaging capacity.

(1) Advancing FRB and pulsar science. The array's high sensitivity provides a major advantage to detect fast radio bursts[26,27], magnetars[28,29], and pulsars[30]. Unlike the VLA or similar arrays, which rely on coherent synthesis of multiple antennas for FRB or pulsar detection sensitivity, the FAST Core Array, with its large, highly sensitive antennas, directly provides sufficient detection sensitivity. Using preliminary detection and arcminute localization of bursts and pulsars with FAST's PAF receivers, targeted antenna correlation processing yielded considerable improvement on the source location precision (at the arcsecond level[31]). By imaging FRB and pulsar environments with unprecedented detail[32], the FAST Core Array will provide essential information on the origin and evolution of FRBs and pulsars.

(2) Resolving the structure of HI galaxies. The strongly enhanced mapping resolution of the FAST Core Array (of the order of arcseconds) relative to FAST (3') will be essential to resolve complex structures in HI galaxies[33–35]. It will allow for the identification of finer structural features, such as galactic disks, tails, bridges, and streams[36,37], and will contribute to the high-precision study of galactic component distributions (including gas, stars, and dark matter)[38–40].

(3) Mapping the cosmic evolution of radio galaxies. Because of its high sensitivity and resolution, the FAST Core Array will detect extremely large populations of radio galaxies[41] and allow for extensive statistical studies of their evolution and morphology as a function of cosmic time[42–44]. Such studies will be essential to address major questions on radio galaxies, for example the influence of galaxy mergers, supernovae, supermassive black holes, and star formation processes on the morphology of radio galaxies at different redshifts[45–47].

(4) Investigating transient phenomena. The FAST Core Array will extend FAST's major contribution to investigations of the transient universe by providing crucial calibration capabilities for follow-up radio observations, flux density measurements, and subarcsecond localization of energetic events such as supernovae, gamma-ray bursts, or neutron stars mergers[48–54]. Thus, the FAST Core Array will improve our understanding of relativistic jets, shockwaves, and powerful outflows associated with these transient phenomena[55,56].

(5) Detecting exoplanetary radio bursts. With its unprecedented sensitivity, the FAST Core Array could detect brown dwarfs or giant exoplanets. This would constitute the first possibility to investigate exoplanetary magnetic fields, rotation rates, star-planet interactions, and habitability with observations acquired at radio wavelengths[57–60].

(6) Identifying tidal disruption events. The FAST Core Array has the potential to detect and monitor the evolution of relativistic jets triggered by tidal disruption events[61–63] when stars approaching a black hole are destroyed by its gravitational force. Tracking jet formation at subparsec scales would provide valuable insights into magnetohydrodynamics and accretion physics near the event horizon[64,65].

(7) Contributing to deep space exploration. In addition to radio astronomy, the FAST Core Array is expected to contribute to deep space science[66]: space situational awareness, detection of small space targets, communication and control of deep space probes, ionospheric measurements, and improvement of pulsar timing standards. It will also contribute to studies of solar system objects and galactic magnetic field mapping using diffuse synchrotron imaging.

## 5. SUMMARY

This article described the FAST Core Array project. It includes the planned integration of the FAST instrument into a synthesis aperture array and comprising 24 secondary 40-m antennas located within 5 km of the FAST site. By combining FAST's exceptional sensitivity with the high resolution provided by the interferometer and by using state-of-the-art technology, notable discoveries are expected for pulsars, FRBs, HI galaxies, exoplanets, and cosmology. Rapid implementation will allow the FAST Core Array to strengthen FAST's major global position among radio astronomy facilities before the FASTA project becomes operational, providing new opportunities to study dynamic phenomena in the universe.

## ACKNOWLEDGEMENTS

This work was supported by the National Key R&D Program of China (2022YFA1602904), the Chinese Academy of Sciences Project for Young Scientists in Basic Research (YSBR-063), and the National Natural Science Foundation of China (12225303 and 12041301).

## AUTHOR CONTRIBUTIONS

Peng Jiang conceived the idea, and played the project administration and supervision role. Rurong Chen wrote the original draft and used software for simulation analysis. Hengqian Gan, Jinghai Sun, Boqin Zhu and Hui Li provided the conceptual design. Weiwei Zhu, Jingwen Wu, Xuelei Chen, Haiyan Zhang and Tao An provided investigation support. Peng Jiang and Tao An reviewed and edited the manuscript. All the authors read and approved the final manuscript.

## DECLARATION OF INTERESTS

Peng Jiang is an editorial board member for Astronomical Techniques and Instruments and was not involved in the editorial review or the decision to publish this article. The authors declare no competing interests.